\documentstyle[aps,multicol]{revtex}

\begin{document}

\author{B.A.Muzykantskii and D.E.Khmelnitskii}

\address{Cavendish Laboratory, University of Cambridge, Cambridge CB3 0HE, UK
\\ and L.D.Landau Institute for Theoretical Physics, Moscow, Russia}

\title{Nearly Localised States in Weakly Disordered Conductors.\\
II. Beyond Diffusion Approximation.}

\date{\it December 27, 1995}

\maketitle

\begin{abstract}
  We use optimal fluctuation method for a new ballistic $\sigma$-model to
  study the long time dispersion of conductance $G(t)$ of a mesoscopic
  sample.  In the long time limit the conductance of a $d$-dimensional
  sample decays as $ \exp \left( -A \ln^d t \right) $. At shorter times the
  new results match those in our previous paper \cite{MK94}. It is found
  that at very long times the diffraction effects are important and the
  ballistic treatment is not valid. We also suggest a physical picture of
  trapping.
\end{abstract} 
 
\begin{multicols}{2}
\narrowtext

\section{Introduction}

    This article is a continuation of our paper \cite{MK94} where we
suggested an optimal fluctuation method to study exponentially rare
fluctuations in weakly disordered conductors. The quantity of interest
was expressed using the super-matrix nonlinear $\sigma$-model \cite{Efetov}, and
the resulting functional integral was evaluated by the saddle point
method. In this way we obtained an intermediate asymptote of conductance
time dispersion in one- and two- dimensional metals.  It was,
however, pointed out that this method has only a limited scope of validity,
e.g., it fails to describe satisfactorily three-dimensional
conductors.  The failure occurs when the super-matrix corresponding to
the optimal fluctuation changes rapidly over the mean free path $l$. Then
the diffusion approximation breaks down and the use of the nonlinear 
$\sigma$-model \cite{Efetov} can no longer be justified.

In this article we treat the problem using the recently proposed
ballistic $\sigma$-model \cite{MK95}: the generalisation of the standard
one that correctly accounts for large gradients. To be specific, we
consider the long time asymptote of conductance dispersion.  If a
voltage $V(t)$ is applied to a sample, the current through it is given by
the Ohm law
\begin{equation}
  \label{respon}
  I(t) = \int_{-\infty}^t  G(t-t') V(t') dt',
\end{equation}
and we are interested in the behaviour of the conductance $G(t)$ in the long
time limit. 

We first discuss this problem under conditions when the diffusion
approximation is valid. The super-matrix theory is defined by the the
partition function (see \cite{Efetov} and \cite {VWZ} for review):
\begin{equation}
  Z=\int
{\cal D} Q e^{-F},\;  F= \frac{ \pi \nu}{8} \mathop{\rm str}\nolimits
  \int d{\bbox r} \{ D ({\bbox \nabla} Q)^2 + 2 i\omega \Lambda Q\},
\label{Efet}
\end{equation}
where the functional integral over super-matrices $Q$ is
subjected to the constraint 
\begin{equation}
Q^2 = 1.
\label{ConstraintQ}
\end{equation}
The expression for the averaged conductance $G(t)$ looks as follows
\begin{equation}
G(t) = G_0 e^{- t/\tau}  +\int\frac{d\omega}{2 \pi} 
e^{-i\omega t} \int\limits_{ Q^2 =1}  {\cal D} Q  P\{Q\} e^{-F}
\label{G}
\end{equation}
In Eqs.~(\ref{Efet}), (\ref{G}) $\nu$ is the density of states, $D$ is
diffusion coefficient and $\tau$ is the mean free time.  The strategy
suggested in Ref \cite{MK94} consists in studying the condition of
the extremum of the free energy $F$ in Eq.~(\ref{Efet})
\begin{equation}
2 D \nabla (Q \nabla Q) + i \omega [\Lambda, Q]=0
\label{Us}
\end{equation}
together with the condition at the boundary $\Gamma$ between   
the mesoscopic sample and a bulk electrode  
\begin{equation}
 Q |_{\Gamma} = \Lambda,
\label{Bound}
\end{equation}
and a self-consistency condition 
 \begin{equation}
\frac{4 t\Delta}{\pi \hbar} = - \int \frac{d {\bbox r}}{V} \mathop{\rm str}\nolimits \{ \Lambda Q
\}; \quad \Delta = \frac{1}{\nu V}.
\label{self-cons-diffusive}
\end{equation}
that arises after integrating out $\omega$ in Eq.~(\ref{G}).
After solving equation (\ref{Us}) with boundary conditions
(\ref{Bound}), expressing frequency $\omega$ through time $t$,  
using self-consistency condition~(\ref{self-cons-diffusive})
and substituting the solution $Q(r)$ to the free energy (\ref{Efet}),
we obtain the conductance $G(t)$ with exponential accuracy.

Long time retardation in electric response $G(t)$ is caused by relatively
improbable quasi-localised states that are weakly coupled to
the bulk electrodes and have life time of the order of $t$. Since the
mean square of the wave function $|\Psi(r)|^2$ is connected to the
super-matrix $Q(r)$ via
\begin{equation}
  \label{psi-and-Q}
   |\Psi(r)|^2 \sim - \mathop{\rm str}\nolimits \{ \Lambda Q  \},
\end{equation}
the self-consistency condition (\ref{self-cons-diffusive}) can be
regarded as a relation between the life-time of a quasi-stationary state
and its wave function. The super-matrix $Q$ is fixed at the boundary of
the sample (see Eq.~(\ref{Bound})), so, to satisfy the self-consistency
condition~(\ref{self-cons-diffusive}) $\mathop{\rm str}\nolimits  \{ \Lambda Q  \}$ must grow
towards the middle of the sample. The gradients of the $Q$-matrix increase
simultaneously with the delay time $t$. Since theory~(\ref{Efet})
correctly accounts only for the lowest term in $\nabla Q$, it cannot be
used for sufficiently long times $t$.

The importance of high gradients for long time asymptotes and tails of
distribution functions was first announced by Altshuler, Kravtsov and
Lerner (AKL) \cite{AKL}, who found the growth of the corresponding
invariant charges under renormalization group flow. The optimal
fluctuation method has been used recently by Falko and Efetov
\cite{EfFal} and by Mirlin \cite{Mirlin1}, who  studied the influence of the
nearly localised states on the distribution of wave function amplitudes
\cite{EfFal} and local density of states \cite{Mirlin1}.  These authors
found that the gradients of $Q$ become large near the centre of the
sample and, therefore, the $\sigma$-model description breaks down.

To treat consistently this problem, along with the others where
ballistic motion of electrons is essential, we suggested a new version of the
nonlinear $\sigma$-model \cite{MK95}. This theory operates with a
super-matrix distribution function $g_{\bbox n} ({\bbox r})$, where ${\bbox n}$ is the unit
vector of the electron momentum direction (${\bbox p} = {\bbox n} p_F$). It effectively
accounts for the infinite series in $l \nabla Q$ of which only the
leading term is kept in the action~(\ref{Efet}), and, therefore,
correctly describes fluctuations of the super-matrix $Q$ with wave vectors $q
\sim 1/l$. The theory is still restricted, however, to the region of
validity of semi-classical approximation. 
The condition of extremum for the new action is related to the 
kinetic equation in the same manner as Eq.~(\ref{Us}) is
related to the diffusion equation.  We solve the kinetic equation for 
one- two- and three-dimensional geometries and obtain very long time
asymptotes of conductance.

The results are summarised in table \ref{table-summary}.  The exponential
decay law for times $ t_D \ll t \ll \hbar/\Delta $ can be obtained
analysing the time dependence of weak localisation corrections.  The
``ballistic'' part of the long time asymptote at $t > t_b$ was first
studied by AKL who used the frequency representation and expanded the
conductance $G(\omega)$ in powers of $\omega$. They estimated the growth
rate of coefficients in this expansion and pointed out the importance of
high gradients.  Unfortunately, the Fourier transformation to
$t$-representation in two-dimensions was carried out with insufficient
precision (see \cite{Mirlin2} for discussion) and the factor $g$ was lost
under the sign of $\ln^2$. The AKL approach also failed to predict the
existence of intermediate asymptote at $\hbar \Delta^{-1} \ll t \ll t_b$
which was discovered by the authors \cite{MK94} using the optimal
fluctuations method for the diffusive $\sigma$-model.  We found the
region of validity for this intermediate asymptote and obtained some
estimates for even longer times.  These estimates were recently improved
by Mirlin~\cite{Mirlin2} who imposed somewhat arbitrary effective
boundary conditions on the super-matrix $Q$ at the point where the
diffusion approximation breaks down. The value of the action turned out
to be rather insensitive to the exact form of these boundary conditions
which enabled Mirlin to rederive the AKL result in 2d and obtain $\exp (-
A \ln^3 t)$ asymptote in 3d, although without the value of the
coefficient $A$ in the exponent.

The full ballistic treatment presented in this article gives the
coefficient in the exponent for 3d; confirms the AKL result in 2d; and
discovers a new regime in the case of a thick wire. We have also found a
restriction on the validity of ballistic treatment. It turns out, that at
ultra-long delay times the super-matrix distribution function depends strongly
on the direction ${\bbox n}$ of the momentum developing sharp features with
characteristic width $ \delta \phi$. At times $t \sim t_Q$ this width
becomes comparable with the diffraction angle $\delta \phi_q \sim
\lambda/a$, where $a$ is a typical size. At longer times the diffraction
effects become important and the ballistic treatment is no longer valid.
The values of times $t_Q$ are presented in Table~\ref{table-summary}.
Apart from derivation of these results from the first principles the
paper describes the optimal fluctuation of random potential in a strictly
one-dimensional wire that traps an electron for time $t$. We believe
that the same mechanism is responsible for long time delays
in higher dimensions.
 
The material is organised as follows.  In section II we present physical
motivations behind the ballistic $\sigma$-model, find its condition of
extremum; and show how the ballistic description transforms into the
diffusive one when the gradients are small. We also discuss the geometry
of the super-matrix distribution function $g_{\bbox n}(r)$ and introduce its
convenient parametrisation. The optimal fluctuation method is
described in section III. In section IV the solution of kinetic
saddle-point equation is found and the long time asymptote of the
conductance is evaluated in two and three dimensions.  The one
dimensional case is discussed in section V.  Physical picture of
trapping is presented in section VI. Finally, in section VII 
we discuss the
results and the limits of their validity.

\section{Effective Action for Quantum Ballistics}

\subsection{Outline of Derivation}

In this section we present a generalised non-linear super-matrix 
$\sigma$-model \cite{MK95}, which is valid in the ballistic regime. 

If the quantum effects are neglected, the ballistic regime 
is described by the Boltzmann kinetic equation 
for the distribution function $f_{\bbox n} ({\bbox r}) $ of 
coordinate $\bf r$ and momentum ${\bf p} = {\bbox n} p_F.$ 
The quantum description operates with density matrix 
$\hat g_{\bbox n} ({\bbox r})$. 
To enable averaging over disorder, $\hat g_{\bbox n} ({\bbox r})$ should be
a super-matrix \cite{MK95}, 
analogously to the matrix $\hat Q$ in the standard $\sigma$-model 
(\ref{Efet}). The quantum generalisation of the kinetic equation has 
the form 
\cite{Eilenberger}  
\begin{equation}
  \label{Eilen}
  2 v {\bbox n} \frac{\partial g_{\bbox n} ({\bbox r})}{\partial {\bbox r}} = 
  \left[ \left( i \omega \Lambda - \frac{\langle g ({\bbox r})\rangle}{\tau} 
  \right),g_{\bbox n} ({\bbox r}) \right]
\end{equation}
with the additional constraint 
\begin{equation}
        g_{\bbox n}^2 = 1,
        \label{Norm}
\end{equation}
which is similar to the one imposed on $Q$.  Equation (\ref{Eilen}) is the
required ballistic generalisation of the saddle-point equation
(\ref{Us}) and serves as an extremum condition for the ballistic
action we are constructing. To generate the first derivative in 
Eq.~(\ref{Eilen}), the action has to have a Wess-Zumino type term
\begin{eqnarray}
  &&{\cal W}\{g_{\bbox n}\} = \int  \int_0^1 du \mathop{\rm str}\nolimits \left \langle \tilde
    g_{\bbox n} ({\bbox r},u) \left[\frac{\partial \tilde g_{\bbox n}}{\partial u}, {\bbox n}
      \frac{\partial \tilde g_{\bbox n}} {\partial {\bbox r}} \right] \right \rangle d {\bbox r}
\label{W}
\\
&&\tilde g_{\bf n} ({\bf r},u=0) = \Lambda;\; 
\qquad \tilde g_{\bf n}({\bf r},u=1) = g_{\bf n}({\bf r}),
\label{Vic}
\end{eqnarray}
where an arbitrary smooth interpolation can be chosen
as  $\tilde {g}_{\bbox n} ({\bbox r},u)$ \cite{Fradkin}, and the angular brackets denote
averaging over directions of ${\bbox n}$.
The functional derivative $\delta {\cal W} / \delta g_{\bbox n} ({\bbox r})$ 
is taken with the constraint (\ref{Norm}) which guarantees that $ g_{\bbox n}
\delta g_{\bbox n} + \delta g_{\bbox n} g_{\bbox n} =0 $ and an arbitrary
variation  $\delta g_{\bbox n}$ has the form
  $\delta g_{\bbox n} = \left[ g_{\bbox n},  a_{\bbox n} \right]$.
As a result of variation 
\begin{equation}
\label{deltaW}
\delta {\cal W} = 4 \int d {\bbox r} \mathop{\rm str}\nolimits \left \langle 
 {\bbox n} \frac{\partial g_{\bbox n}}{\partial {\bbox r}} a_{\bbox n} 
 \right \rangle
\end{equation}
the first derivative appears.There is another way of writing the
functional $\cal W$ which employs the decomposition $ g_{\bbox n} = U \Lambda
U^{-1}$:
\begin{equation}
  \label{W-vs-U}
  {\cal W}\{g_{\bbox n}\} = 4 \int d {\bbox r}   \mathop{\rm str}\nolimits \langle \Lambda U^{-1} {\bbox n}
  \frac{\partial U}{\partial {\bbox r}} \rangle.
\end{equation}
This representation can be verified by comparing the variation of $\cal
W$ with Eq.~(\ref{deltaW}).
The quantum ballistic partition function $Z$ can be presented as an
integral over distribution functions $g_{\bbox n} ({\bbox r})$ with effective action
$F$:
\begin{mathletters} 
\label{BigVic}
\begin{eqnarray}
 Z&=&\int\limits_{g_{\bbox n}^2=1} \!\!\! {\cal D} g_{\bbox n} 
({\bbox r}) e^{-F}, 
\label{BigVic1} \\ 
 F &=& \frac{\pi
    \nu}{4}  \int d {\bbox r}  \mathop{\rm str}\nolimits \left\{
    i\omega \Lambda \langle g ({\bbox r}) \rangle  - \frac{1}{2 \tau} 
\langle g({\bbox r})\rangle^2 \right\} -
\nonumber \\
&& - \frac{\pi \nu v_F}{8} {\cal W} \{ g_{\bbox n} \}.
\label{BigVic2}
\end{eqnarray}
\end{mathletters}
The details of the derivation can be found in Ref \cite{MK95} . 

Field theory (\ref{BigVic}) enables us to study any chaotic problem, for
which the semi-classical approach is valid, irrespective of validity of
the diffusion approximation.  If space gradients are small ($ l \nabla g
\ll g $), the standard treatment recovers (see \cite{MK95}) the
$Q$-matrix theory (\ref{Efet}).  To show this we expand the matrix
$g_{\bbox n}$ into a sum over angular harmonics keeping only the zeroth and
first terms:
\begin{equation}
g_{\bbox n} = Q({\bbox r}) \left( 1 - \frac{Q {\bbox J}^2}{2}\langle {\bbox n}^2 \rangle \right) + 
{\bbox J} ({\bbox r}) \cdot {\bbox n}
        \label{subst}
\end{equation}
The constraint $g^2 = 1$ now reads
\begin{equation}
        Q^2 =1, \qquad Q {\bbox J} + {\bbox J} Q = 0. 
 \label{constr7}
\end{equation}
Substituting Eq.~(\ref {subst}) into Eqs.~(\ref {BigVic}) and using
conditions (\ref{constr7}), we obtain the partition function in the
form
\begin{eqnarray}
  Z&=& \int {\cal D} Q \int {\cal D} {\bbox J} e^{-F_{Q, {\bbox J}}}, 
\nonumber \\
F_{Q, {\bbox J}} &=&
  \frac{\pi \nu}{4} \int d {\bbox r} \mathop{\rm str}\nolimits \{ i\omega \Lambda Q +
  \frac{{\bbox J}^2}{6\tau } - \frac{v_F}{3} ({\bbox \nabla} Q) Q {\bbox J} \}
        \label{poldorogi}
\end{eqnarray}
The Gaussian integral over ${\bbox J}$ in Eq.~(\ref {poldorogi}) is dominated 
by the vicinity of 
\begin{equation}
  \label{current}
  {\bbox J} = l ({\bbox \nabla} Q ) Q
\end{equation}
and  leads finally to Eq.~(\ref {Efet}). 

\subsection{Symmetries of $g$-matrices} 
Both $Q$ and $\Lambda$  are  $ 8 \times 8$
matrices which act on the $8$-component super-vectors $\Psi$ with the basis
\cite{Efetov}:
\begin{equation}
\Psi^\top = (\chi_1, \chi_1^*, S_1, S_1^*, \chi_2, \chi_2^*, S_2,
S_2^*),  
\label{basis-efetov}
\end{equation}
where $\chi$ are the fermionic and $S$ are bosonic variables; 
indices $1,2$ correspond to the retarded and advanced Green
functions with energies $E_f \pm \omega/2$.  For a super-matrix
$\hat{M}$ the super-trace is defined as follows:
\begin{eqnarray*}
\mathop{\rm str}\nolimits \hat{M} &=& M_{11} + M_{22} -M_{33} -M_{44} + \\
                   & & M_{55} + M_{66} -M_{77} -M_{88} 
\end{eqnarray*}

In this basis the super-matrix $\Lambda$ has the form
$$ 
\Lambda_{ij} = \lambda_i \delta_{ij}, \quad
\lambda_i = \left \{ \begin{array}{rr} 1,\quad & i=1 \ldots 4 \\
                                     -1,\quad & i=5 \ldots 8
                   \end{array} \right.
$$
For our purposes it is more convenient to use another basis where each 
variable is stands next to its charge conjugate partner and
fermionic variables are separated from bosonic: 
\begin{equation}
 \Psi^\top = (\chi_1, \chi_2^*,\chi_2, \chi_1^*,  S_1, S_2^*, S_2,
S_1^*).
\label{Abasis}  
\end{equation}
In this new basis, which will be used everywhere in this paper,
the matrix  $\Lambda$ has the form:
\begin{equation}
  \label{Lambda}
  \Lambda =\left(
    \begin{array}{cc}
      \sigma_3 \otimes \tau_3   & 0\\
      0&   \sigma_3 \otimes \tau_3
    \end{array}
\right),
\end{equation}
where the blocks in Eq.~(\ref{Lambda}) correspond to the fermionic and
bosonic sectors.  The matrices ${\bbox \tau}$ act inside the $2\times 2$
blocks, while matrices the ${\bbox \sigma}$ mix these blocks inside the $4\times 4$
sectors.

We first discuss the symmetry of the Q-matrix from the standard
$\sigma$-model (\ref{Efet})  focusing on the properties of
the boson-boson ($B$) and fermion-fermion ($F$) sectors only. We consider oly the
unitary ensemble when an additional condition
\begin{equation}
  \label{unitary}
  \left[ Q^{B,F}, 1 \otimes \tau_3 \right] = 0
\end{equation}
is fulfilled. Together   with  constraint (\ref{ConstraintQ}),
requirement (\ref{unitary}) permits to parametrise the matrices
$Q^{B,F}$ with four complex 3-vectors ${\bbox l}^{B,F}, {\bbox m}^{B,F}$
subject to the constraint ${\bbox l}^2={\bbox m}^2=1$:
\begin{equation}
  \label{param-Q0}
  Q^{B,F}= \frac{1 + \tau_3}{2} \otimes {\bbox l} {\bbox \sigma} + \frac{1 -
    \tau_3}{2} \otimes {\bbox m} {\bbox \sigma}.
\end{equation}
Parametrisation (\ref{param-Q0}) is still too general because the matrix
$Q$ has additional symmetries: charge neutrality
\begin{mathletters}
  \label{symm}
\begin{equation}
  \bar Q \equiv CQ^\top C^\top = Q,  
  \label{symm-charge}
\end{equation}
with the charge conjugation matrix
$$C=\left( 
\begin{array}
{cc}
   i \sigma_2 \otimes \tau_1 & 0 \\
   0 & -\sigma_2 \otimes \tau_2
\end{array}
     \right),
$$
and pseudo-hermiticity
\begin{equation}
  Q^{\dagger} \equiv K Q^+ K = Q, \quad 
C=\left(
\begin{array}
{cc}
   1 & 0 \\
   0 & \sigma_3 \otimes \tau_3
\end{array}
     \right).
  \label{symm-hermit}
\end{equation}
\end{mathletters}
Requirements (\ref{symm-charge}) are satisfied when
\begin{mathletters}
  \label{l-and-m}
\begin{eqnarray}
&& {\bbox l}^B = - {\bbox m}^B, \qquad {\bbox l}^F = - {\bbox m}^F, \\
\hbox{and} \nonumber \\
&& {\bbox l}^B=\hat \mu ({\bbox l}^B)^*,\; {\bbox l}^F= ({\bbox l}^F)^*, \;
\hat \mu = \left(\begin{array}[c]{ccc}
-1 & 0 & 0 \\
0 & -1 & 0 \\
0 & 0 & 1
\end{array}\right).
\end{eqnarray}
\end{mathletters}
Thus, the fermionic sector is parametrised by a real vector ${\bbox l}^F$
subjected to constraint $({\bbox l}^F)^2=1$, i.e. a sphere ${\cal S}^2$ 
\begin{equation}
  {\cal S}^2 = \left\{ {\bbox l}^F,\quad l_1^2 + l_2^2 + l_3^2 =1,\quad
    \mbox{Im } {\bbox l} =0\right\}
\label{sphere}
\end{equation}
while the bosonic sector is represented by the vector ${\bbox l}^B$ with two
imaginary components $l_1^B$ and $l_2^B$ and one real $l_3^B$.  Due to
constraint ${\bbox l}^B=1$ the bosonic sector is represented by a hyperboloid
${\cal H}_2^2$:
\begin{equation}
\label{hyperb}
{\cal H}_2^2 = \left\{ {\bbox l}^B, \quad -|l_1|^2 - |l_2|^2 + l_3^2 =1,\quad
  {\bbox l} = \hat \mu {\bbox l}^* \right\}.
\end{equation}

The matrix $g_{\bbox n}$ of the ballistic $\sigma$-model (\ref{BigVic}) obeys the
constraint $g_{\bbox n}^2 =1$ and, therefore, can be parametrised with
Eq.~(\ref{param-Q0}).  The condition of charge neutrality \cite{Comment} 
(\ref{symm-charge}) must be replaced by the generalised version
\begin{equation}
  \label{symm-g}
  \bar g_{\bbox n} \equiv Cg_{\bbox n}^\top C^\top = g_{-{\bbox n}}
\end{equation}
because the charge conjugation changes the sign of the $\cal W$-term in
action (\ref{BigVic}). 

The pseudo-hermitian transformation 
$g_{\bbox n} \to K g_{\bbox n}^+ K$ not only changes the sign of the
$\cal W$-term but also replaces $\omega$ by $-\omega^*$. In the long time
asymptote calculation the frequency $\omega$ is purely imaginary and
the generalisation of Eq.~(\ref{symm-hermit}) reads:
\begin{equation}
 K g_{\bbox n}^+ K  = g_{-{\bbox n}}
\label{herm-g}
\end{equation}

Symmetries (\ref{symm-g}) and (\ref{herm-g}) impose a  new restriction
on the vectors ${\bbox l}$ and ${\bbox m}$ (compare with Eqs.~(\ref{l-and-m})):
\begin{equation}
  \label{l-and-m-g}
   {\bbox l}_{\bbox n}^B = - {\bbox m}_{-{\bbox n}}^B, \; {\bbox l}^F = - {\bbox m}_{-{\bbox n}}^F, \;
   {\bbox l}_{-{\bbox n}}^B = \hat \mu ({\bbox l}_{\bbox n}^B)^*, \; {\bbox l}_{-{\bbox n}}^F = ({\bbox l}_{\bbox n}^F)^*.
\end{equation}
Since conditions (\ref{l-and-m-g}) are imposed on the componets of two
vectors ${\bbox l}$ and ${\bbox m}$, they are less restrictive than
Eqs.~(\ref{symm}).  Both $g_{\bbox n}^B$ and $g_{\bbox n}^F$ can be parametrised
with a complex unit vector ${\bbox l}_{\bbox n}= {\bbox \xi}_{\bbox n} + i{\bbox \eta}_{\bbox n}$:
\begin{equation}
  \label{param-g}
  g_{\bbox n}^{B,F}= \frac{1 + \tau_3}{2} \otimes {\bbox \sigma} {\bbox l}_{\bbox n} + \frac{1 -
    \tau_3}{2} \otimes {\bbox \sigma} {\bbox l}_{-{\bbox n}}.
\end{equation}
with the constraint
\begin {equation}
{\bbox l}^2=  {\bbox \xi}^2 -{\bbox \eta}^2 + 2 i ({\bbox \xi} {\bbox \eta}) =1.
\label{xieta}
\end{equation}
The geometric meaning of Eq.~(\ref{xieta}) can be described as follows.
The vector ${\bbox \xi} = {\bbox \nu} \xi$ is characterised by its absolute value $\xi$
and a unit vector ${\bbox \nu}$, which corresponds to a point on a sphere ${\cal
  S}^2$. Due to the condition $0={\bbox \xi} {\bbox \eta} = \xi {\bbox \nu} {\bbox \eta}$, the vector
${\bbox \eta}$ belongs to the plane tangential to the sphere $ {\cal S}^2$ at
the point ${\bbox \nu}$. The condition $\xi^2 - {\bbox \eta}^2 = 1$ means that at every
point of the sphere there is an upper part of the two sheet hyperboloid
${\cal H}_2^2$ with the axis along the radius of the sphere.
In other words, the vector ${\bbox l}$ subject to constraint (\ref{xieta}),
belongs to the fibre bundle with the base $ {\cal S}^2$ and the fibre 
${\cal H}_2^2$. The $Q$-matrices in this geometric picture are represented
by the sub-manifold of this fibre bundle: $Q^F$ belongs to the base and
$Q^B$ lies on the fibre over the North Pole of ${\cal S}^2$ (see
Eq.~(\ref{sphere})).  The $\Lambda$-matrix corresponds to  the bottom
of  ${\cal H}_2^2$ in the fibre over the North Pole.

\section{steepest descent procedure for quantum ballistics}

The long time asymptote of the conductance $G(t)$ is found following
the procedure outlined in the introduction. The generalised version 
of Eq.~(\ref{G}) has the form 
\begin{equation}
G(t) = G_0 e^{-t/\tau}
+\int\frac{d\omega}{2\pi}e^{-i\omega t} \int_{g^2=1} {\cal D} g_{\bbox n} 
P\{g_{\bbox n}\} e^{-F},
\label{Gt2}
\end{equation}
where $F$ is the ballistic action (\ref{BigVic2}), and the explicit form of
the functional $P$ is irrelevant within  exponential accuracy. The
functional integral (\ref{Gt2}) over $g_{\bbox n}$ is evaluated using the
steepest descent method, which consists in:
\begin{enumerate}
\item finding a solution $ g_n(r)$ of the saddle point equation
  (\ref{Eilen});

\item expressing the frequency $\omega$ through the time $t$ using the
  self-consistency condition
  \begin{equation}
    \label{self-cons-ballisitcs}
    \frac{4 t\Delta}{\pi \hbar} = - \int \frac{d {\bbox r}}{V} \mathop{\rm
      str}\nolimits \{ \Lambda \langle g \rangle \},
  \end{equation}
  which arises as a result of integration over $\omega$ in
  Eq.~(\ref{Gt2});
\item substituting $g_n(r)$ in Eq.~(\ref{Gt2}) and obtaining the result
  with exponential accuracy.
\end{enumerate}

We do not specify boundary conditions for Eq.~(\ref{Eilen}) bearing in
mind that far from the centre of the sample the space gradients become
small and $g_{\bbox n}(r)$ approaches a solution of diffusion equation
(\ref{Us}). It was shown in Ref.\cite{MK94} that for the long times $ t
\gg \hbar /\Delta$ the solutions of diffusion equation (\ref{Us}) can be
written in the form:
\begin{equation}
  \label{Q-solution}
  Q = U  \left(
  \begin{array}{cc}
 \sigma_3  \otimes \tau_3 & 0 \\
0 & (\sigma_3 \cosh \theta + i \sigma_2 \sinh \theta ) \otimes \tau_3
  \end{array} \right) U^{-1},
\end{equation}
where the real angle $\theta$ obeys the equation 
\begin{equation}
  \label{theta}
  D \nabla^2 \theta + i \omega \sinh \theta =0; \quad
\theta|_{\mbox{lead}}=0
\end{equation}
and the constant super-matrices $U$ commute with $\Lambda$. They do
not enter the action and thus will be omitted. As can be seen from
Eq.~(\ref{Q-solution}), only the bosonic block $Q^B$ has a non-trivial
dependence on the coordinates. The same holds true in the ballistic region.
and from now we consider only the bosonic sector of the theory. 

Combining the diffusion asymptote (\ref{Q-solution}) with
Eqs.~(\ref{subst},\ref{current}, \ref{param-g}), we find that  the vector 
${\bbox l}_{\bbox n}$  that parametrises $g_n$ approaches the limit
\begin{equation}
  \label{l-diffusive-limit}
    l_1= -l {\bbox n} {\bbox \nabla} \theta, \quad l_2 = i \sinh \theta, \quad l_3 =
  \cosh \theta
\end{equation} 
far from the centre of the sample. One can see from 
Eq.~(\ref{l-diffusive-limit}) that
\begin{equation}
  \label{imaginary-and-real}
    \mbox{Im } l_1 = \mbox{Re } l_2 = \mbox{Im } l_3 = 0
\end{equation}
It turns out that the condition (\ref{imaginary-and-real}) remains valid
even in the ballistic regime.  Therefore, for the solution of kinetic
equation (\ref{Eilen}) the complex vector ${\bbox l}_n(r)$ is restricted to the
real three dimensional subspace (\ref{imaginary-and-real}). The
additional constraint ${\bbox l}^2=1$ thus acquires the form
\begin{equation}
  \label{1-sheet-hyper}
         l_1^2 -|l_2|^2 + l_3^2 =1 
\end{equation}
and defines a one sheet hyperboloid. Convenient coordinates on the subspace
(\ref{imaginary-and-real}) are associated with the cone $ l_1^2 -|l_2|^2
+ l_3^2 =0$ and described in appendix \ref{appendix-conica}.

Due to the symmetry (\ref{l-and-m-g}), $l_1$ is an odd and $l_{2,3}$ are even
functions of ${\bbox n}$.  Averaging Eq.~(\ref{Eilen}) over directions of ${\bbox n}$
and using the parametrisation (\ref{param-g}), we arrive at the continuity
condition
\begin{equation}
2 v_F \hbox{div}\langle {\bbox n} l_1 \rangle + \omega \langle l_2 \rangle= 0
\label{discontin}
\end{equation}
which reduces to the conservation law for the current:
\begin{equation}
{\bbox J} \equiv \langle {\bbox n} l_1 \rangle, \qquad \mbox{div} {\bbox J} =0, 
\label{conserv1}
\end{equation}
if the frequency is small and $\gamma \equiv i \omega \tau \rightarrow 0$. 

\section{Solution of Kinetic Equation and Long time Asymptote of 
conductance}
\label{section-23}

\begin{figure}[p]
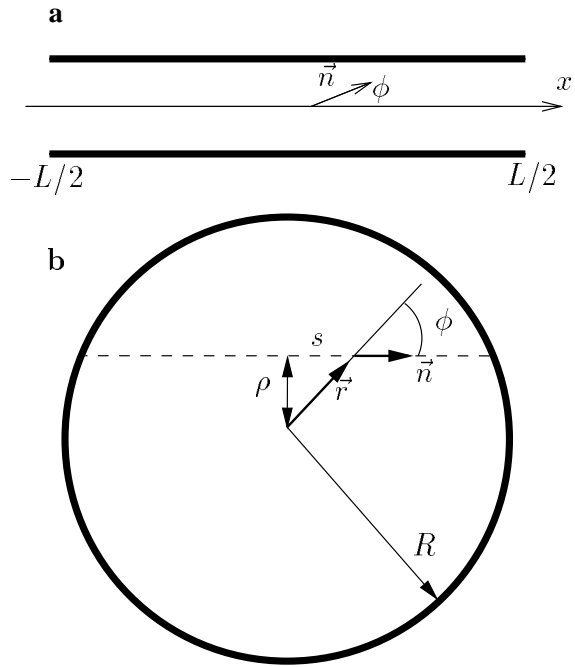

\caption[]{\label{fig-samples-geom} The geometry of the 1d sample (a) 
and 2d and 3d samples (b)}
\end{figure}

In Ref \cite{MK94} we considered the solution of the diffusion equation
(\ref{theta}) either for a disordered wire (1D), or for a disc (2D), or
for a droplet (3D) of radius $R$ (see Fig \ref{fig-samples-geom}) .  In
analysing the kinetic equation (\ref{Eilen}) we stick to the same
geometry. In this section we find the solutions of Eq.~(\ref{Eilen}) in
space dimensions 2 and 3, postponing the discussion of one-dimensional
case to the next section.

Long times correspond to small frequencies $\omega$ ($\gamma \ll 1$).
This means that the terms containing $\gamma$ in the kinetic equation can
be neglected everywhere, except the central part of the sample.  It can
be seen from Eq.~(\ref{conserv1}) that the total flux of the current
${\bbox J}$ is conserved in the outer area. Since the current ${\bbox J}$ is directed
along the radius, flux conservation means that in two- and
three-dimensional samples the current density decays towards the outer
boundary together with the space gradients of all relevant quantities,
and the solution of the kinetic equation approaches the diffusion
asymptote. The situation is, however, different in the one-dimensional
case when the current and gradients do not decay and the diffusion regime
is reached only outside the sample.

\subsection{Qualitative description}

In the space dimensions 2 and 3 the qualitative behaviour of the
solutions is the same for both the diffusive and kinetic equation.  We
illustrate it with the diffusion Equation (\ref{theta}) which can be
regarded as describing a chemical reaction where $\theta$ is the
concentration of a reacting agent. The term $ D {\bbox \nabla}^2 \theta$
describes the propagation of the agent in a porous medium and the rate of
the agent reproduction depends on its concentration as $\gamma \sinh
\theta$.  Since $\gamma \ll 1$, generation takes place only in the
central part of the sample, where the value of $\theta$ is high. We refer
to this central zone as  the ``zone of reaction''.  In the outer part of the
sample the agent diffuses freely (the ``run-out zone''), i.e. $\theta$ is
a solution of the Laplace equation ${\bbox \nabla}^2 \theta =0$.  An azimuthally
symmetric solution $\theta(r)$ obeying the boundary condition
$\theta(R)=0$ decays like
\begin{mathletters}
\label{theta-at-large-r}
\begin{eqnarray}
  \theta = C \ln \frac{R}{r} &\quad& (\hbox{2D}) \\
  \theta = C \left( \frac{1}{r} - \frac{1}{R} \right) &\quad& (\hbox{3D}) 
\end{eqnarray}
\end{mathletters}
as $ r \to R$. 

The separation into two zones is also valid for solutions of
the kinetic equation. It turns out that in the run-out zone the space
gradients are small and the diffusion asymptote (\ref{l-diffusive-limit})
with $\theta$ given by (\ref{theta-at-large-r}) is reached.

In the centre of the sample ${\bbox l}_{\bbox n}$ does not depend on ${\bbox n}$, because the
solution is azimuthally symmetric. Taking into account that $l_1$ is an
odd function of ${\bbox n}$ and using the constraint (\ref{1-sheet-hyper}), we
can define the value of ${\bbox l}_{\bbox n}$ at $r=0$ with a single parameter
$\theta_0$:
\begin{equation}
  \label{l-at-r=0}
l_1(0)=0, \quad l_2(0)= i \sinh \theta_0, \quad l_3(0) = \cosh \theta_0.
\end{equation}
In the next subsection we find the exact solution of the kinetic equation
for $\theta_0= \ln(1/\gamma)$. Although it does not match the asymptote
(\ref{l-diffusive-limit}), (\ref{theta-at-large-r}) at large $r$, it
plays an important role.  Any solution of the kinetic equation that
starts at $\theta_0 < \ln(1/\gamma)$ eventually approaches the diffusion
asymptote, while the one that starts at $\theta_0 > \ln(1/\gamma)$ does not.
Therefore this exact solution is a separatrix.

\begin{figure}[p]
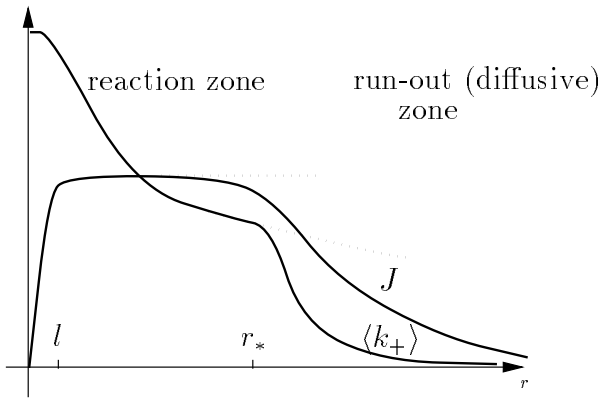

\caption[]{\label{fig-separatrix} Sketch of kinetic equation solutions in
  two- and three- dimensional samples. (dotted line is the separatrix solution)}
\end{figure}

Now we are ready to describe the shape of the solution that obeys the
boundary conditions (\ref{l-at-r=0}) and
reaches the diffusion asymptote at large $r$.  If
$\gamma \ll 1$, it starts at $\theta_0$, being only slightly smaller than
$ \ln 1/\gamma$, and, therefore, runs close to the separatrix up to the
large radius $r_* \gg l$. For even larger $r$ the deviation becomes
significant and our solution crosses over to the diffusion asymptote
(\ref{l-diffusive-limit}) with $\theta$ given by
(\ref{theta-at-large-r}).  The cross-over region is of the order of the
mean free path $l$ and is much smaller than both the reaction and run-out
zones. We match the separatrix in the reaction zone with the diffusion
asymptote in the run-out zone keeping the mean value $\langle {\bbox l}
\rangle$ continuous at the point $r_*$ (see Fig.~\ref{fig-separatrix}), thus
finding both the position of the cross-over $r_*$ and the coefficient $C$
in Eq.~(\ref{theta-at-large-r}).

\subsection{Solutions}
An azimuthally symmetric solution of the kinetic equation depends only on the
radius $r$ and the angle $\phi$ between the radius and the vector ${\bbox n}$.
Therefore, the space derivative in the kinetic equation has the form
\begin{equation}
  \label{nnabla-vs-r-theta}
  {\bbox n} {\bbox \nabla} = \cos \phi \frac{ \partial }{\partial r} - \frac{\sin
    \phi}{r} \frac{\partial}{\partial \phi} = \left( \frac{ \partial
      }{\partial s} \right)_\rho,
\end{equation}
where the impact parameter $ \rho = r \sin \phi$ and the distance along
a straight line trajectory $s = r \cos \phi$ are introduced (see
Fig.~\ref{fig-samples-geom}). 

If the parameterisation (\ref{param-g}) for the matrix $g_{\bbox n}$ is used
together with the conic basis for the vector ${\bbox l}_{\bbox n}$ (see
appendix~\ref{appendix-conica}):
 \begin{equation}
  \label{l-vs-k}
  l_1 = k_1, \quad l_3 - i l_2 = k_+ + \gamma, \quad l_3 + i l_2 = k_- +
  \gamma,
\end{equation}
the kinetic equation is simplified to the form (in this section we measure
all distances in the units of the mean free path $l$):
  \begin{mathletters}
\label{Eilen-k}
\begin{eqnarray}
 && \frac{\partial}{\partial s} k_+ = - \langle k_+ \rangle k_1, \\ 
 && \frac{\partial}{\partial s} k_- = \langle k_- \rangle k_1,
\label{k-}\\
&& k_1^2 + (k_+ + \gamma) (k_-+ \gamma)=1.
\label{k-constraint}
\end{eqnarray}
\end{mathletters}
The condition at the origin (\ref{l-at-r=0}) has the form:
\begin{equation}
  \label{k-at-r=0}
  k_1(0)=0, \quad k_+ (0) = e^{\theta_0} - \gamma, \quad k_- (0) = 
  e^{-\theta_0} - \gamma.
\end{equation}
The separatrix solution starts at $\theta_0= \ln ( 1/\gamma ) $ and therefore,
$k_-(0)=0$.  Equation (\ref{k-}) for $k_-$ now gives $k_-=0$ for all $r$.
Taking into account the strong inequality $\gamma \ll 1$, we simplify the
system (\ref{Eilen-k}):
\begin{mathletters}
\begin{eqnarray}
&& \frac{\partial k_+}{\partial s} = - \langle k_+ \rangle k_1,
\\
&& k_1^2 + \gamma k_+ =1
\label{inden2}
\end{eqnarray}
\end{mathletters}
and obtain a closed integro-differential equation for $k_+$:
\begin{equation}
  \label{k}
  \frac{\partial k_+}{\partial s} = - \langle k_+ \rangle \sqrt{1 -
    \gamma k_+}
\end{equation}
The solution of this equation is given in 
appendix~\ref{appendix-solution}; at distances $r \ge 1$ it has the form:
\begin{mathletters}
\label{separatrix-solution}
\begin{eqnarray}
\label{separatrix-solution-k_+}
 \gamma k_+ &&=  1 - \frac{a^2}{4} \ln^2 \cot
    \frac{\phi}{2}  + \frac{ab}{\gamma r}  \frac{(\frac{\pi}{2} -
      |\phi|)\ln \cot \frac{\phi}{2}}{|\sin\phi|} \\ 
\label{separatrix-solution-a}
  a &=& 2 \left \langle \ln^2 \cot 
        \frac{\phi}{2} \right\rangle ^{-1/2} = \frac{4}{\pi} \left\{ \begin{array}{ll}
1  & \quad (2D) \\
 \sqrt{3} & \quad (3D)
\end{array}
\right. \\
b&=&\left\langle \frac{(\frac{\pi}{2} -|\phi|) \ln \cot \frac{\phi}{2}}
{|\sin\phi|} \right\rangle^{-1} = \left\{
\begin{array}{cl}
4 \ln^2 r &\quad (2D)\\
\approx .95 &\quad (3D)
\end{array}
\right.
\end{eqnarray}
\end{mathletters}
The divergence of $k_+$ at $\phi = 0$ is cutoff at the the angles $\phi
\sim 1/r^2$. Since all integrals with $k_+$ converge, the exact form of
this cut-off is relevant only for the criterium of validity of the ballistic
treatment (see section VII). The firstot term in (
\ref{separatrix-solution-k_+}) has zero average and does not depend on
$r$, while the mean value $\langle k_+ \rangle = a/(\gamma r) $ decays
when $r \to \infty$. Using Eq.~(\ref{inden2}), we find the expressions
for the component $k_1$ and the current $J = \langle \cos \phi k_1
\rangle$
\begin{mathletters}
\begin{eqnarray}
  \label{separatrix-solution-k1}
  k_1 &=& - \frac{a}{2} \ln |\cot \frac{\phi}{2}| \\
  \label{separatrix-solution-J}
  J &=&\frac{a}{2} \left\langle \cos \phi \ln |\cot \frac{\phi}{2}| \right\rangle =
  \frac{1}{\pi} \left\{
\begin{array}{cl}
2 &\quad (2D)\\
\sqrt{3} &\quad (3D)
\end{array}
\right.
\end{eqnarray}
\end{mathletters}
Note, that the the current $J$ don't depend on the coordinates.

We are looking at a solution of Eq.~(\ref{Eilen-k}) that transfers to the
diffusion regime at a certain radius $r_* \gg 1$. The transition region
has a width of the order of unity, and, therefore, both the mean value
$\langle k_+ \rangle $ and the current $J$ change negligiblely across the
cross-over region. Combining Eqs.~(\ref{l-diffusive-limit}),
(\ref{theta-at-large-r}) and (\ref{l-vs-k}) we obtain in the diffusion region:
\begin{mathletters}
\label{solution-at-large-r}
\begin{eqnarray}
  \label{k-at-large-r}
  \langle k_+ \rangle = \left(\frac{R}{r}\right)^{C}, \quad J = 
   \frac{Cl }{2r} &\quad& (2D) \\ 
  \langle k_+ \rangle = \exp \frac{C}{r}, \quad J =  \frac{Cl}{r^2}
  &\quad& (3D). 
\end{eqnarray}
\end{mathletters}
Using continuity of $J$ and $\langle k_+ \rangle$ at the
point $r_*$ we find the values of the parameters $ C, r_*$:
\begin{equation}
\label{23d-parameters}
\begin{array}{lll}
         r_* = \frac{\pi l}{4} \frac{\ln \frac{1}{\gamma}}{\ln 
         \frac{R}{l}},& \quad C = \frac{\ln \frac{1}{\gamma}}{\ln 
         \frac{R}{l}}& \quad (2D),\\
         r_* = \frac{\pi l}{3\sqrt{3}} \ln \frac{1}{\gamma}, 
         &\quad C =\frac{\pi l}{3\sqrt{3}} \ln^2 \frac{1}{\gamma}
         & \quad (3D)
\end{array}
\end{equation}
The sketch of the solution is given on the Fig.~\ref{fig-separatrix}.

\subsection{Long-time Asymptote of Conductance}

We begin with the self-consistency condition (\ref{self-cons-ballisitcs}).
In the conic coordinates (\ref{l-vs-k}) it looks as follows:
\begin{equation}
  \label{self-cons-conica}
  \frac{\Delta t}{ \pi \hbar} =  \int \frac{d {\bbox r}}{V} \frac{ \langle k_+ -
  k_- \rangle +2\gamma}{2}\approx \int \frac{d {\bbox r}}{V} \frac{\langle k_+ \rangle}{2},
\end{equation}
where it is taken into account that for solution
(\ref{separatrix-solution}) the inequality $k_- \ll k_+$ holds.  The last
integral can be calculated using the continuity condition
(\ref{discontin}), where $-i l_2= (k_+ - k_-)/2 $ can
be replaced by $k_+/2$. Integrating Eq.~(\ref{discontin}), we express the
integral (\ref{self-cons-conica}) through the total current $J$ through the outer
boundary of the sample:
\begin{equation}
  \label{self-cons=total-flux}
  \int_{r=R} {\bbox J}(r) dS = \frac{\gamma}{2l} \int d {\bbox r} \langle k_+ \rangle.
\end{equation}
Although the integral (\ref{self-cons-conica}) contains comparable
contributions from both the reaction and run-out zones, with the help of
Eq.~(\ref{self-cons=total-flux}) it can be expressed through the current in the
run-out zone only. Using the asymptote (\ref{solution-at-large-r}) we finally
arrive at the expression for $\gamma$:
\begin{mathletters}
  \label{gamma-vs-t}
\begin{eqnarray}
  \label{gamma-vs-t-2d}
  \gamma = \frac{\pi}{2} p_F l \frac{ \tau}{t} \ln \left(\frac{t}{p_F l
      \tau}\right), \quad (2D),\\
  \label{gamma-vs-t-3d}  
  \gamma = \frac{2\pi}{\sqrt{3}} (p_F l)^2 \frac{\tau}{t} \ln^2
  \left(\frac{t}{\tau (p_F l)^2} \right), \quad (3D)
\end{eqnarray} 
\end{mathletters}
The long time asymptote of the conductance $G(t)$ is determined by the
value of the action $F$ from Eq.~(\ref{BigVic2}) for solution
(\ref{separatrix-solution}) - ( \ref{23d-parameters}).  This action is a
sum of two contributions: from the run-out zone ($F_D$) and from the
reaction zone ($F_b$). The action is dominated by $F_D$:
\begin{eqnarray}
  \label{action-D}
  F_D &=& \frac{\pi \nu D}{2} \int_{r_*}^R d {\bbox r} \left(\frac{d \theta}{d r}
  \right)^2 = 
\nonumber \\
&& \pi^2 \nu D C^2  \left\{ 
    \begin{array}[c]{lcl}
       \ln ( R / r_* ) & \quad & (2D) \\
      2 / r_* & \quad & (3D) 
    \end{array} \right.
\end{eqnarray}
Substituting the values of the parameters (\ref{23d-parameters}) we obtain:
\begin{eqnarray}
\label{final-action-2D}
F_D = \frac{\pi g}{2} \frac{ \ln^2 t/(\tau g) }{ \ln R/l} &\quad &
      (2D)\\
\label{final-action-3D}
F_D = \frac{ \pi}{9 \sqrt{3}} ( p_F l)^2 \ln^3 \frac{t}{\tau g}& \quad & (3D).
\end{eqnarray}
where $g= 2\pi \hbar\nu D$ is the dimensionless conductance. 
The contribution from the
reaction zone $F_b$ is calculated, using ballistic action (\ref{BigVic2}),
 in appendix~\ref{appendix-action-1d}: 
\begin{mathletters}
  \label{action-b}
\begin{eqnarray}
  F_b \sim 
  g \frac{\ln (t / g \tau}{\ln(R/l)} &\quad & (2D) 
\\ 
  F_b \sim 
(p_F l)^2 \ln^2   \left( \frac{t}{(p_F l)^2 \tau} \right) &\quad& (3D)
\end{eqnarray}
\end{mathletters}
 Comparing Eqs.~(\ref{final-action-2D}), (\ref{final-action-3D}) and
(\ref{action-b}) we see that the action is dominated by the contribution
from the diffusion zone. In two-dimensions it comes from the whole
diffusion zone $r_* \ll r \ll R$ because the integral in Eq.~(\ref{action-D})
diverges logarithmically. In three-dimensions only the region near
$r=r_*$ of width of the order of $r_*$ is important.
Since $r_* \gg l$, the contribution from the crossover region, which has 
the width of the order of $l$, can be neglected. The long time
conductance asymptote is given by
\begin{equation}
  \label{answer-23d}
  G(t)= \exp \left( -F_D \right)
\end{equation}
with the action $F_D$ defined Eqs.~(\ref{final-action-2D}) and 
(\ref{final-action-3D}).

\section{Long-time asymptote for Conductance in Disordered Wire}

In this section we consider a 1D thick wire of length $L$ and
cross-section $w$ ( $ w \ll l \ll L$) with specular boundary
conditions and assume that the distribution function $g_{\bbox n}$ is uniform
across the wire.  For not very long times the diffusion equation
(\ref{theta}) is valid and has the solution 
\begin{equation}
  \label{theta-at-large-r-1d}
  \theta= \theta_0 - |\frac{x}{\xi}|, \quad \xi= L/\log{t \Delta}.
\end{equation}
The space gradient $\nabla \theta = 1/\xi $ does not depend on the
coordinate $x$ and is smaller than $1/l$ as long as the time is shorter
than $\hbar \Delta^{-1} \exp (L/l)$. For longer times the diffusion
regime breaks down simultaneously in the whole wire.  The separation into
reaction and run-out zone is still valid, and the term in 
$\omega$ in the kinetic equation is important in the reaction zone near
the centre of the wire and can be neglected elsewhere. Unlike the 2D and
3D cases however, the space gradients in the run-out zone do not decay
towards the outer ends of the wire. We now present a solution of the
kinetic equation in the run-out zone which is valid for arbitrary
gradients.

The distribution function $g_{\bbox n}$ depends on the coordinate $x$ along the
wire and the angle $\phi$ between the axis of the wire and the direction ${\bbox n}$
of the momentum (see Fig.~\ref{fig-samples-geom}). In these coordinates 
the kinetic equation (\ref{Eilen}) takes the form:
\begin{equation}
  \label{Eilen-1d}
  2 l \cos \phi \frac{\partial g}{\partial x} = \left[ \left( 
      \gamma \Lambda - \langle g \rangle \right),g \right]
\end{equation}
In the run-out zone the term in $\gamma$ is negligible and, using the
conic coordinates, from appendix~\ref{appendix-conica} as $\gamma
\rightarrow 0$, we rewrite Eq.~(\ref{Eilen-1d}):
\begin{mathletters}
\label{Eilen-conica-1d}
\begin{eqnarray}
  \label{Eilen-conica-1d-k}
  l \cos \phi \frac{\partial}{\partial x} k_\pm = \mp \langle k_\pm
  \rangle k_1, \\
  \label{Eilen-conica-1d-const}
  k_1^2 + k_+ k_-=1.
\end{eqnarray}
\end{mathletters}
Dividing both sides of Eq.~(\ref{Eilen-conica-1d-k}) by $\cos \phi$ and
averaging over $\phi$, we find a closed equation for $ \langle k_\pm
\rangle$ with the solutions:
\begin{equation}
  \label{1d-k-pm}
  \langle k_\pm \rangle = q \exp ( \mp \theta),
\end{equation}
where $q$ is a constant and $\theta$ obeys the equation:
\begin{equation}
  \label{1d-theta}
  l \frac{d \theta}{dx}= \langle \frac{k_1}{\cos \phi} \rangle.
\end{equation}
In one-dimension the current conservation law~(\ref{conserv1}) reduces to $J
\equiv \langle k_1 \cos \phi \rangle = \mbox{const}$ and suggests that
$k_1$ does not depend on $x$. Therefore,  Eq.~(\ref{1d-theta})
gives
\begin{equation}
  \label{1d-theta-2}
   l \frac{d \theta}{dx}= \hbox{const} \equiv \theta'
\end{equation}
and, substituting this back into Eq.~(\ref{Eilen-conica-1d-k}), we obtain the
solution in the factorised form:
\begin{equation}
  \label{1d-factor}
  k_\pm = \beta(\phi) \langle k_\pm \rangle, \quad  k_1 = \theta' \beta(\phi)
  \cos \phi, 
\end{equation}
where the function $\beta(\phi)$ is determined from
Eq.~(\ref{Eilen-conica-1d-const}) 
\begin{equation}
\label{1d-beta}
\quad \beta =
  [q^2 + (\theta')^2 \cos^2 \phi]^{-1/2} 
\end{equation}
Finally, the constants $\theta'$ and $q$ are related by the condition
\begin{equation}
  \label{1d-j0-and-q}
  1 = \langle \beta \rangle = \frac{1}{2\pi} \int_0^{2\pi} 
  \frac{d \phi}{\sqrt{q^2 +  (\theta')^2 \cos^2 \phi}}
\end{equation}
Using the asymptotic values of the elliptic integral in
Eq.~(\ref{1d-j0-and-q}), we find
\begin{equation}
\label{1d-theta-q }
q = \left\{
\begin{array}[c]{cl}
1 -\frac{\theta'^2}{4}, &\qquad  \theta' \ll 1,\\
4 \theta' \exp \Big( -\frac{\pi |\theta'|}{2} \Big) &\qquad \theta' \gg 1
\end{array}
\right.
\end{equation}
and obtain expressions for the current $J= \langle k_1 \cos \phi \rangle$:
\begin{equation}
\label{1d-J}
J = \left\{
\begin{array}[c]{cl}
\frac{\theta'}{2}, & \quad  \theta' \ll 1,\\
\frac{2}{\pi} \left\{ 1 - 4 \pi|\theta'| \exp \Big( -\pi
      |\theta'| \Big)\right \} & \quad \theta' \gg 1
\end{array}
\right.
\end{equation}

The above solution is not valid in the reaction zone in the vicinity of
$x=0$, whose contribution to both the action and self-consistency condition
is negligible. The boundary condition for the ballistic problem is
determined by the requirement that the distribution function matches the
one in the bulk electrodes where $g_{\bbox n}=\Lambda$.  For small gradients
($\theta' \ll 1$) the angular dependence of the distribution function in
the wire is weak and the above condition is fulfilled provided
$\theta(\pm L/2)=0$. On the other hand, in the ballistic regime ($\theta'
\gg 1$) the distribution function in the wire strongly depends on the
angle $\phi$ (see Eq.~(\ref{1d-beta})). Thus, there is a cross-over
region near the outer ends of the wire where the solution of the kinetic
equation deviates from Eqs.~(\ref{1d-k-pm}), (\ref{1d-factor}) and
(\ref{1d-beta}).  We consider the wires which are long enough to neglect
the change of $\theta$ in the cross-over region.  Therefore, the
boundary condition $\theta(\pm L/2)=0$ is still valid in the ballistic
regime, and the function $\theta(x)$ can be presented in the form
(\ref{theta-at-large-r-1d}) with $\xi=l/\theta'$ and $\theta_0= \theta'
L/(2l)$.

\subsection{Action and self-consistency condition}
The self-consistency condition (\ref{self-cons-conica}) gives
\begin{equation}
  \frac{t\Delta}{\pi \hbar} \frac{L}{l} =\frac{1}{|\theta'|} 
\exp \left(\frac{ |\theta'| L}{2l}\right)
\label{1d-self-cons}
\end{equation}
and should be compared with the continuity equation
(\ref{self-cons=total-flux})
 which in the 1D case reads:
\begin{equation}
  J = \frac{\gamma}{2|\theta'|} \exp \left(- \frac{ |\theta'| L}{2l}\right).
\label{1d-contin}
\end{equation}
Using Eq.~(\ref{1d-J}) for the current and
Eqs.~(\ref{1d-self-cons}), and  (\ref{1d-contin}) we
express $\gamma$ and $\theta'$ through $t$:
\begin{mathletters}
\label{1d-gamma-theta}
\begin{eqnarray}
\label{1d-theta'}
|\theta'|&=&\frac{2l}{L} \cdot \Big(\ln\frac{t\Delta}{\pi\hbar} 
+ \ln \ln \frac{t\Delta}{\pi\hbar} \Big),\\
\label{1d-gamma}
\gamma &=& \frac{2 \pi \hbar}{t\Delta} \frac{l}{L} J = \frac{\tau}{t}
\left\{
\begin{array}{cl}
2 g \ln \frac{t\Delta}{\pi\hbar} & \quad \frac{t\Delta}{\pi\hbar} \ll 
\exp\Big(\frac{L}{l}\Big),\\
2 N &\quad \frac{t\Delta}{\pi\hbar} \gg \exp\Big(\frac{L}{l}\Big)
\end{array}
\right.
\end{eqnarray}
\end{mathletters}
where $g= 2\pi \hbar \nu D w/ L$ is dimensionless conductance and $N = w
p_F/\pi \hbar$ is the number of transverse channels, which is equal to
ballistic conductance.  The calculation of the ${\cal W}$-term in
action~(\ref{BigVic2}) for the solution (\ref{1d-factor}) is performed in
appendix~\ref{appendix-action-1d} and gives:
\begin{equation}
  \label{action-W-1d}
 {\cal W}\{g_{\bbox n}\} = -8 L w \frac{\theta' J}{l}
\end{equation}
Evaluating the other terms in the action we get:
\begin{equation}
        F = \frac{\pi}{2 \Delta \tau}(2J \theta'+ q^2 -1).
        \label{action-1d}
\end{equation}
Thus, the long time asymptote of the conductance is given by:
\begin{equation}
\label{1d-answer} 
\begin{array}[c]{lc}
G(t) \sim 
  \exp\left\{ -g \ln^2 \left( \frac{t\Delta}{2\pi \hbar} \right) \right\} &
   1 \ll \frac{t\Delta}{2\pi \hbar} \ll
  e^{L/l} \\ 
G(t) \sim \left( \frac{t \Delta}{2\pi \hbar}
  \right)^{-2N}  & e^{L/l}\ll
  \frac{t\Delta}{2\pi \hbar} \ll N e^{L/l}
\end{array}
\end{equation}
The last inequality in Eq.~(\ref{1d-answer}) ensures the validity of
semiclassical approximation (see section VII).

\section{physical picture of trapping}
In this section we consider the time dispersion of the conductance $G(t)$
in a purely one-dimensional wire. This problem was solved by Altshuler and
Prigodin \cite{Altshuler-Prigodin} who obtained the long time asymptote
in the form
\begin{equation}
  \label{pure-1d-conductance}
  G(t) \sim \exp \left( - \frac{l}{L} \ln^2 t \Delta \right),  
\end{equation}
which can be treated as a limiting case of a multi-channel formula for a
thick wire (see table~\ref{table-summary}), assuming that in a
one-channel case $g$ is given by $l/L$.
Formula~(\ref{pure-1d-conductance}) can be understood as a probability of
an optimal potential fluctuation that traps an electron of Fermi energy
$E_f$ for time $t$. In a weak potential $ U(x) \ll E_f$ the wave function
can be presented in the form
\begin{equation}
  \label{psi-1d}
  \Psi(x) = \phi_+(x) e^{i p_F x} + \phi_-(x) e^{-i p_F x}
\end{equation}
with the amplitudes $\phi_{\pm}(x)$ changing slowly: $\nabla \phi_\pm \ll
p_F \phi$. Let us consider a quasi-stationary state obeying the open
boundary conditions
\begin{equation}
  \label{psi-boundary}
  \phi_+(0)=\phi_-(L)=0,
\end{equation}
which correspond to the outward flow of current through the ends
of the wire.  The life time of such a state is inversely proportional to
the outward current
\begin{equation}
  \label{psi-lifetime}
  t= \frac{\int dx |\Psi|^2 }{ v_F \left( |\phi_-(0)|^2 +
      |\phi_+(L)|^2 \right)}.
\end{equation}
The maximum delay time is achieved when the currents through 
both ends are equal
($\phi_-(-L/2)= \phi_+(L/2)$). Fixing normalisation by
\begin{equation}
  \label{psi-norm}
  |\phi_-(0)|= |\phi_+(L)| =1
\end{equation}
we reduce Eq.~(\ref{psi-lifetime}) to the from
\begin{equation}
  \label{pis-lifetime-2}
\frac{t \Delta}{\pi \hbar} = \int \frac{dx}{L} |\Psi|^2 
\end{equation}
which resembles the self-consistency condition~(\ref{self-cons-diffusive})  
obtained for arbitrary dimensions.

To achieve life times $t \gg \hbar/\Delta$ the wave function must grow
towards the middle of the wire; assuming the growth is exponential,
$ \Psi \sim \exp[(L/2 - |x|)/\xi]$, we obtain for the localisation 
length of the quasi-stationary state
\begin{equation}
  \label{psi-xi}
  \xi = \frac{L}{ \ln t \Delta}.
\end{equation}
A typical random potential $\tilde{U}$ causes one-dimensional wave
functions to be localised with $\xi \sim l$. The shorter localisation
length $ \xi \ll l$ corresponds to life times longer than $\hbar
\Delta^{-1} \exp(L/l)$ and can be achieved in the potential
\begin{equation}
  \label{psi-U}
  U(x) = \tilde{U} + U_0 \cos (2 p_F x).
\end{equation}
with the additional $2 p_F$-Fourier component having the amplitude
\begin{equation}
  \label{psi-amplitude}
  U_0 = \frac{2 \hbar v_F }{\xi}.
\end{equation}
The probability of this potential realization in  given by the 
Gaussian distribution:
\begin{equation}
  \label{psi-probability}
  \exp - \left( \frac{\pi \nu \tau}{2} \int U(x)^2 dx \right) \sim \exp
      \left( - \frac{l}{L} \ln^2 t \Delta \right).
\end{equation}
and coincides with~(\ref{pure-1d-conductance}) with a correct numerical
factor  in the exponent. We therefore
conclude that the states with long life times are locked by the Bragg
reflection and can be found with the probability proportional to that of
potential fluctuation with the Bragg mirror of appropriate strength.

We believe that the same mechanism is responsible for nearly localised
states in multi-channel wires and the samples of higher dimensionals. In
multi-channal cases, however, adding a single $2p_F$-Fourier harmonics
cannot localise the wave function, because the random part $\tilde U$
mixes different directions of the momentum. To localise the wave function in a
two- or three- dimensional sample, the potential fluctuation should be
effective for all directions of the momentum; so we expect it to have the form
\begin{equation}
  \label{psi-multi-u}
  U(x) = \int d \Omega_{\bbox n}  U_{\bbox n}({\bbox r}) \cos ( 2 p_F {\bbox n} {\bbox r} )
\end{equation}
with the amplitude $U_{\bbox n}({\bbox r})$ slowly depending on ${\bbox n}$ and ${\bbox r}$.

\section{discussion}
This paper continues our study of the conductance $G (t)$ asymptote at
long times $t$ that began in Ref.~\cite{MK94}. We use the steepest
descent approach which enables us to obtain $g(t)$ for different ranges
of time. The purpose of this section is dual: we want to analyse the
restrictions of our treatment and compare our results with those in the
literature.

Let us first discuss the conductance $G(t)$ of a thick wire made from a
two-dimensional strip of length $L \gg l$ and width $w < l \ll L$.  The
solution $g(x,\phi)$ of the  kinetic equation has typical
gradients (see Eq.~(\ref{action-W-1d})):
\begin{equation}
  \label{discussion-gradients}
  \frac{\partial g}{\partial x} \sim \frac{\theta'}{l} \sim \frac{\ln (t
    \Delta/\hbar)}{L}
\end{equation}
At long times the distribution function changes rapidly with the angle
$\phi$ acquiring a sharp maximum at $\phi= \pi/2$ with the width $ \delta
\phi \sim \exp (- \pi |\theta'|/4)$ (see
Eq.~(\ref{1d-beta})-(\ref{1d-J})). The semi-classical approximation remains
valid if this width is much larger than the diffraction angle $\delta
\phi_Q \sim p_F w$. This imposes an upper limit on the gradients:
\begin{equation}
  \label{discussion-theta'}
  \theta' \ll \ln (p_F w /\hbar)
\end{equation}
Thus, the ballistic asymptote is valid only for times smaller than
\begin{equation}
  \label{discussion-time-Q}
  t_Q \sim \Delta^{-1} p_F w e^{L/l}
\end{equation}
Since the width of the wire is limited by the condition~\cite{wide-wire}
$w <l$ the gradient $\theta'$ at times $t \sim t_Q$ is still much smaller
than $p_F$.  Thus, an intermediate interval of gradients arises
\begin{equation}
  \label{discussion-theta-tQ}
   \ln (l p_F ) \ll   \theta' \ll  l p_F 
\end{equation}
which corresponds to the interval of delay times:
\begin{equation}
  \label{discussion-tQ}
  \frac{L}{l} \ln \frac{l}{\lambda} \ll \ln (t \Delta) \ll
  \frac{L}{\lambda}.
\end{equation}
We assume that the time dispersion of the conductance in this region can
be recovered by some kind of a ballistic treatment with diffrction proper
accounted for.

A similar phenomenon restricts the range of applicability of ballistic
asymptotes in two- and three- dimensional samples. Analogously to the
one-dimensional case, the distribution function $g(\phi,r)$ in the
reaction zone has a sharp feature at $\phi=0$ with the width
$\delta\phi=l/r$. Diffraction can be neglected when $\delta\phi \gg
\lambda/l $. Therefore, the reaction zone radius $r_*$ must obey the
constraint:
\begin{equation}
  \label{discusion-2d}
  r_* \ll l^2/\lambda.
\end{equation}
Substituting the radius of reaction zone from Eq.~(\ref{23d-parameters}),
we obtain the upper time limit $t$ for  which the semi-classical approach is
still valid:
\begin{equation}
  \label{discusion-t-2d}
  t \ll t_Q \sim \hbar \Delta^{-1} \exp \left( \frac{p_f l}{\hbar}
  \right) \left\{ \begin{array}{cl} l/R & \quad (2D) \\ 
                   (l/R)^3 &\quad   (3D)
                  \end{array} \right.
\end{equation}
Similarly to the one-dimensional case the gradients of $g$ at the time
$t_Q$ are still much smaller than $p_F$, being only of the order of
$1/l$.  The liming times $t_Q$ for a different geometries are summarised
in table \ref{table-summary}.

As it has been mentioned in the introduction, the whole field was
pioneered by AKL. They added high powers of  frequency and gradients of
$Q$ to the diffusive $\sigma$-model and analysed the renormalization flow
of the corresponding coupling constants in two dimensions. This gave the
growth rate of the coefficients $C_n$ in the expansion 
\begin{equation}
  \label{discussion-G(w)}
  G(\omega)= \sum_{n=0}^\infty C_n \omega^n.
\end{equation}
The Fourier transform of Eq.~(\ref{discussion-G(w)}) with the AKL asymptote
for $C_n$ leads to the result presented in Table~\ref{table-summary}. AKL
also put forward a general conjecture  that the logarithmically normal
asymptote was valid for all dimensions.

As one can see from table~\ref{table-summary}, there is a variety of
different regimes in the time dispersion of the conductance. The AKL
procedure predicts one of them and fails to describe the others. It is also
difficult, using this procedure, to find out the criteria when this or
that results are valid.

Another attack on the problem has been recently carried out by Mirlin
\cite{Mirlin2}. He used our optimal fluctuation method for the diffusive
$\sigma$-model in dimensions two and three and supplied Eq.~(\ref{Us}) by
somewhat arbitrary the conditions near the origin.  In this way he obtained
the asymptotes of conductance for 2D and 3D samples. The reason for his
success lies in the fact established in this paper that the action for
the dimensions $d \ge 2$ is dominated by the contribution from the run-out
zone.  The solution of the ballistic problem confirms the result of
Ref.\cite{Mirlin2} for two-dimensional conductance. In the three-dimensional
case we obtain the numerical coefficient $A$ in the expression
$G(t) \sim \exp(-A \ln^3 t)$. Although the form of the density-matrix
fluctuation suggested by Mirlin deviates strongly in the reaction zone
from the correct one, the method of his work could be useful for
qualitative estimates.

It should be noted, however, that the ballistic treatment is needed to
find the upper limit $t_Q$ on the delay time interval where these results
are valid. We regard the occurance of the the new range of times $t \ge
t_Q$ where the diffraction effects are important we regard as one of the
most interesting results of this paper.

\section{Acknowledgement}

We thank A.D Mirlin for sending us his paper \cite{Mirlin2} prior to
publication and B.I.Shklovskii for an inspiring discussion.

\appendix
\section{conic coordinates}
\label{appendix-conica}
Throughout this paper we use a convenient parametrisation of the bosonic
sector of the distribution function $g_{\bbox n}$. To obtain it, we introduce
a basis in the space of traceless $2\times 2$ real matrices:
\begin{equation}
\hat e_1 = \sigma_1, \qquad \hat e_{\pm}=\frac{\sigma_3 \pm i \sigma_2}{2}.
\label{e}
\end{equation}
The new unit vectors have the following properties:
\begin{mathletters}
\label{e-ort}
\begin{eqnarray}
&&\hat e_1^2 = 1,  \; \hat e_{\pm}^2 = 0, \;
\left[ \hat e_{\pm} , \hat e_1 \right] = \pm 2 \hat e_{\pm},
\label{e-ort-1} \\
&&[\hat e_+ , \hat e_- ] = - \hat e_1 ,\;  \{\hat e_+ , \hat e_- \}= 1.
\label{e-ort-2}
\end{eqnarray}
\end{mathletters}
Now, instead of parametrisation~(\ref{param-g}) for the $g^B_{\bbox n}$, we use:
\begin{eqnarray}
  \label{conica-param-g}
&&  g_{\bbox n}^B = \frac{1 + \tau_3}{2} \otimes \left( l_1({\bbox n}) \hat e_1 + l_+({\bbox n})
  \hat e_+ + l_-({\bbox n}) \hat e_- \right) + \nonumber \\ 
&& + \frac{1 - \tau_3}{2}
  \otimes \left( l_1(-{\bbox n}) \hat e_1 + l_+(-{\bbox n}) \hat e_+ + l_-(-{\bbox n}) \hat
    e_- \right),
\end{eqnarray}
where, due to requirement (\ref{imaginary-and-real}), all components $l_1$
and $l_{\pm}$ are real. The kinetic equation can be further simplified by
introducing the functions
\begin{equation}
  \label{condica-k}
  k_1 = l_1; \quad k_+ = l_+ - \gamma; \quad k_- = l_- - \gamma
\end{equation}
The constraint $g^2=1$ now can be written as:
\begin{equation}
  \label{conica-constraint}
  k_1^2 + (k_+ +\gamma)(k_- + \gamma)=1,
\end{equation}
and the kinetic equation~(\ref{Eilen}) in
coordinates~(\ref{nnabla-vs-r-theta}) has the form:
\begin{mathletters}
\label{conica-Eilen}
\begin{eqnarray}
 && \frac{\partial}{\partial s} k_+ =- \langle k_+ \rangle k_1, \\ 
 && \frac{\partial}{\partial s} k_- = \langle k_- \rangle k_1, \\
 && \frac{\partial k_1}{\partial s} = -k_+ \langle k_- \rangle + 
        \langle k_+ \rangle k_- +\gamma \langle k_+ - k_-\rangle.
        \label{conica-k_1}
\end{eqnarray}
\end{mathletters}
It is more convenient to use the constraint (\ref{conica-constraint})
instead of the last equation in this set (see Eq.~(\ref{Eilen-k}) in the text).

\section{integro-differential equation for separatrix}
\label{appendix-solution}
Solving Eq.~(\ref{k}) with respect to $k_+$ and considering its average
value $\langle k_+ \rangle$ as a given function of radius $r$, we obtain:
\begin{equation}
  1- \gamma k_+ = \frac{\gamma^2}{4} \left( \int_0^s \langle k_+ \rangle
    ds'\right)^2.
        \label{sol}
\end{equation}
By taking the average of both the right- and left-hand sides, this
expression is reduced to an integral equation for $\lambda(r) = \gamma
\langle k_+ \rangle$:
\begin{equation}
\label{Int}
 1 -\lambda (r) = \frac{1}{4}\left\langle \left( \int_0^{r\cos \phi} ds 
\lambda\left(\sqrt{r^2 \sin^2 \phi + s^2 }\right)
\right)^2\right\rangle_{\phi},
\end{equation} 
where
$$
\begin{array}{cl}
\langle \rangle_{\phi} =
\int_0^{2\pi} \frac{d\phi}{2\pi} &\quad (2D) \\ 
\langle \rangle_{\phi} =
\int_0^{\pi} \frac{\sin \phi d\phi}{2} &\quad (3D)
\end{array}
$$
The solution of Eqs.~(\ref{Int}) has the following asymptotes:
\begin{equation}
\label{Int2}
\lambda (0) = 1; \qquad \lambda(r) = \frac{a}{r} + \frac{b}{r^2} 
,\; \; \; r \gg 1,
\end{equation}
where the constants $a$ and $b$ are given by 
Eqs.~(\ref{separatrix-solution}) in the text.

If $r\sin \phi \gg 1$, then the asymptote of (\ref{Int2}) can be substituted
into kernel of Eq.~(\ref{sol}), and give the result
\begin{equation}
\label{B4}
\gamma k_+ = 1 - \frac{a^2}{4} \ln^2 \left(\cot\frac{\phi}{2} \right) +
\frac{ab}{2r}\frac{\ln \left((\frac{\pi}{2}-|\phi|) |\cot
    \frac{\phi}{2}|\right)}{|\sin \phi|} 
\end{equation}
Expression (\ref{B4}) is valid only if $r|\sin\phi|\gg 1$ and is free of
singularities in this region.

\section{Calculation of Ballistic Action in  the reaction zone}
 \label{appendix-action-2d}

The contribution from the reaction zone to the action is given by
Eq.~(\ref{BigVic2}) and contains two terms:
\begin{equation}
  \label{appendix-2,3action}
  F_1 = - \frac{\pi \nu v_F}{8} {\cal W}, \quad
  F_2 = - \frac{\pi\nu}{8 \tau} \int  \langle g \rangle^2;
\end{equation}
the term with $\omega$ dissapears after integration over $\omega$
in Eq.~(\ref{Gt2}).

To caluclate the ${\cal W}$ term we use Eq.~(\ref{W-vs-U}) and present
the solution of kinetic equation in the form $g_{\bbox n} = U \Lambda U^{-1}$.
Using parametrisation (\ref{param-g}) we find for the bosonic block of
$g$-matrix:
\begin{equation}
  \label{appendix-g-2,3d}
  g^B = \left( 
  \begin{array}{cc} 
    U(\phi) \sigma_z U^{-1}(\phi)& 0 \\ 0 & - U(\pi -\phi) \sigma_z
    U^{-1}(\pi -\phi)
  \end{array} \right),
\end{equation}
where $2\times2$ matrix $U$ is given by:
\begin{equation}
  \label{appendix-u-2,3d}
    U= \exp \left(- \frac{k_1}{\gamma} \hat e_+ \right) 
     \exp \left( \frac{\ln \gamma}{2} \sigma_1 \right).
\end{equation}
Using 
$$ 
\frac{ \partial}{\partial r} U = -\frac{1}{\gamma}\frac{ \partial
  k_1}{\partial r} \hat e_+ U
$$ 
we obtain from Eq.~(\ref{W-vs-U}):
\begin{equation}
  \label{appendix-W-2,3}
  {\cal W} = 8 \int d {\bbox r} \langle {\bbox n} \frac{ \partial k_1}{\partial
      r} \rangle = 8 \int d S {\bbox J},
\end{equation}
where the integral in the last expression is taken over the boundary of
the reaction zone.
This gives
\begin{equation}
  \label{appendix-F1-2,3}
  F_1 \sim \left\{
\begin{array}{cl}
g \frac{\ln (t / g \tau}{\ln(R/l)} &\quad (2D) \\
(p_F l)^2 \ln^2 \left( \frac{t}{(p_F l)^2 \tau} \right) &\quad (3D)
\end{array} \right.
\end{equation}
which is one power of the logarithm smaller than the contribution from
the run-out zone.

The calculation of the  term $F_2$ is strait-forward:
\begin{equation}
  \label{appendix-F2-2,3}
  F_2 =  - \frac{\pi\nu}{8 \tau} a^2 \int d {\bbox r} \frac{l^2}{r^2} \sim \left\{
  \begin{array}{cl}
    g \ln \left( \frac{\ln (t / (g \tau)}{\ln(R/l)} \right) &\; (2D) \\
    (p_F l)^2 \ln \left( \frac{t}{(p_F l)^2 \tau} \right) &\; (3D)
  \end{array} \right.
\end{equation}

\section{Calculation of  $\cal W$ term for one-dimensional solution.}
\label{appendix-action-1d}

To employ the expression (\ref{W-vs-U}) for the $\cal W$-term we should 
find a decomposition $g_{\bbox n} = U \Lambda U^{-1}$. 
Combining Eqs.~(\ref{1d-k-pm}) and (\ref{1d-factor}) with the 
parametrisation (\ref{param-g}) and formulae from
appendix~\ref{appendix-conica}, we write the bosonic block of $g_{\bbox n}$ in
the form of (\ref{appendix-g-2,3d}), where
the  $2\times2$ matrix $U$ is given by:
\begin{equation}
  \label{appendix-U}
  U= \exp \left(\frac{\theta}{2} \sigma_x \right) 
     \exp \left(- i \frac{\chi}{2} \sigma_y \right),
\end{equation}
where  $\chi = \mbox{ arcsin } \left( \theta' \beta(\phi) \cos \phi
\right)$ does not depend on $x$.
The straight-forward application of Eq.~(\ref{W-vs-U}) gives:
\begin{equation}
  \label{appendix-W-1d}
  {\cal W} = -8 L w  \frac{|\theta'| J}{l}
\end{equation} 
where the minus sign comes from the  super-trace of the bosonic block.
The result of averaging over ${\bbox n}$ is expressed  through the
current $J= \langle \cos \phi k_1 \rangle$.

\end{multicols}

\widetext

\begin{table}[p]
$$
\begin{array}{|c|c|c|c|c|c|}
\hline
& t_D \ll t \ll \hbar/\Delta  & \hbar / \Delta \ll t \ll t_b 
& t_b \Delta &  t_b \ll t \ll t_Q & t_Q \Delta
\\
\hline
\mbox{1D}\tablenote{strip of width $w$ and lenth $L$,
$N= w p_F/(\pi \hbar)$ is the number of ballisitc channels} & 
 t/t_D     & g \ln^2(t \Delta)   &  \exp (L/l)  & 
2 N \ln t \Delta & N \exp(L/l)
\\
\mbox{2D}\tablenote{disk of radius $R$} & 
\pi t /t_D & 4 g \ln (t\Delta)  & (L/l)^2 &
 \pi g \cdot \frac{\ln^2(t/g\tau)}{2 \ln (R/l)} & 
\exp (p_F l/\hbar) \frac{l}{R}
\\
\mbox{3D}\tablenote{ball of radius $R$} & 
\pi t /t_D & \mbox{ none }  & 1  &
 \frac{ \pi  }{9 \sqrt{3}} (p_F l)^2 \ln^3 
\left[ \frac{t}{\tau (p_F l)^2 }\right] &
\exp (p_F l/\hbar) \left(\frac{l}{R}\right)^3
\\
\hline
\end{array}
$$
\caption[Long time asymptotes of the conductance.]{\label{table-summary}
  Long time asymptotes of the conductance $G(t)$. The cells contain
  formulae for $-\ln G$ at different time intervals for one- two- and
  three- dimesional samples. The diffusion time $t_D$ is defined as
  $L^2/D$ (1d) and $R^2/D$ (2d,3d); the diffusion coefficient $D$ is
  given by $\frac{1}{2} v_F^2 \tau $ (1d, 2d) and $\frac{1}{3} v_F^2 \tau
  $ (3d); $\tau$ is the mean free time and $\Delta=(\nu V)^{-1}$ is the
  mean level spacing.}
\end{table}

\newpage
\narrowtext

\begin{thebibliography}{11}
\bibitem{Comment} The symmetry of the solutions of the kinetic
  equation~(\ref{Eilen}) differs from the symmetry of $g$-matrices in the
  functional integral (\ref{BigVic1}), i.e. the integration countour
  should be deformed to reach the saddle point. Analogous situation
  occurs in the diffusive $\sigma$-model. In the text we describe the
  symmetries of the solutions of the kinetic equation.
\bibitem{wide-wire} If the wire is wider than $l$ the distribution function depends
    on the transverse coordinate and the problem should be treated as a
    two-dimensional one
\end{thebibliography}
\end{document}